\documentstyle{article}
\begin{document}
\begin{center}
{ 8 levels  of harmony and 8 concepts  of Complex Systems
}\\
{D.B. Saakian}\\

\end{center}
\begin{abstract}
A set of general physical principles is proposed as the structural basis for
the theory of complex systems. First the concept of harmony is analyzed and its 
different aspects are uncovered. Then the concept of reflection is defined and illustrated 
by suggestive examples. Later we propose the principle of (random) projection of
symmetrically expanded prereality as the main description method of complex systems.

 \end{abstract}
 
To understand complex phenomena [1] we suggest to detalize the concepts
of harmony and reflection [2].\\
{\bf 1. Harmony.}

Considering Random Energy Model (REM)[3] in physics , swan neck as a symbol of beauty and concept of harmonic person I have found 8 levels of harmony. First 6 levels are general, the latter ones are specific for alive systems.\\
1. Symmetry, global or local.\\
 Local symmetry could be considered as information processing property.\\
2. Variation principle.\\
3. Parametric resonance or Nishimori temperature [4] effect. \\If there 
is a hierarchy in a system and it is possible to define the essence for
each of its levels (a word, a number or a sign), there is a harmony,
if they coincide. For the case of reflection one can define a harmonic reflection when the essence of reflected reality coincides with the essence of reflection.\\
4  Multi-logical reading of a  system.\\
Every deep truth has several faces and any interesting
physical system allows different ways of solution-REM, Hydrogen atom, 2d Ising model....\\
5. Edge of chaos or existence of almost opposite pure qualities
in the same system.\\
Such situation was essential for evolution [5],[6], d=1 barrier in strings
also belongs to this case [7],[2].\\
6. Modalities. \\
R.S.Ingarden,A. Kossakowski,M. Ohya introduced [8] modalities as 
"possible non- categorical attitudes to reality". From this point of view space, potential 
energy, classical entropy, classical information, quantum entropy   and
information are steps of the hierarchical staircase. I add to this list the number of replicas .
Perhaps there are some harmonic situations here: "golden section", two replicas as multicritical
  point in generalization of [9]...\\
 One can distinguish different modalities by the level of complexity, so it will be reasonable 
 to expect, {\bf the more is complex the system, the higher it should climb on the modality staircase.}\\ 
7.Principle of purity.\\
 It is connected with the points 2. and 5., but it is something more.
The survival of system (vital energy!) is connected with the degree in which system can reveal almost opposite
pure properties. This property is crucial for human, without it any serious harmony
 is impossible.  It will be interesting to define it on quantitative level for other complex 
 systems. Here  it could be connected with conservation or circulation of some free energy
  among the hierarchy levels and while  losing this property system  becomes dissipate.  \\
8.Principle of minimal reflections.\\
This principle explicitly works in ethics (nobody likes words "not","but" and likes modesty) and economics
 (too much stocks create crush). 

{\bf 2. Reflection}

Reflection [2] is a mapping from the original space to a mapped space which\\
1.Touches all the parts of system;\\
2.Conserves certain crucial  interrelations  of original system;\\
3.Compresses phase space or changes the total direction.

If it is possible to define free energy connected with the reflection, then the
situation is interesting. 
From the point of view of the observer the original  phase space is connected 
with objective reality, and  subjective reality- with the reflected 
phase space (magnetizations, temperature, pressure...). 
When there is some  free energy, logarithmic
from the number of degrees (schemata of Gellmann), we define it as complex system.
According to this classification critical theories, 2d quantum disorder, multiscaling are complex.

The concepts of harmony and  (logarithmic free energy) reflection have a good 
applications in ethics and aesthetics. If in physics the "crucial  interrelations  within the  original system"
is the manageable amount of motion: energy at micro-level,  free energy at macro-level then 
 Harmony and Reflection are crucial 
for drastically simplification of the picture of the  universe. An observer has a small 
complexity, he can accept small amount of information. There are a lot of hierarchy levels, 
but not only we can survive now but also people in primordial societies have survived just due to this 
property (harmonic reflection): every hierarchy level
 carries essence of the previous (deeper one), so every one can understand the truth after it 
 is simplified in a proper way (proper times of reflection). 
The number of hierarchy levels or reflections of a complex system which are 
harmonic, is the main 
characteristics of any complex system.\\
{\bf 3. Local symmetry losing.}\\
For many interesting situations in physics a system having  local symmetry properties in a normal phase,
is then partially  loosing it while becoming complex [2],[10].\\
{\bf 4. Transformation of subjective reality into objective one.}\\
If different level of hierarchies are well separated, in principle, we can miss the low 
ones. The most complex situation corresponds to the birth of a new reality. This corresponds
to the edge of chaos situation or the border of errorless decoding in information theory
(the border of true and false)[10].\\
   {\bf 5. Projection of symmetrically(canonically, unitarily or replicaly) expanded  prereality.}\\
 While one has been considering together different forms of (matter) motions , theory 
became contradictory. To overcome the difficulties
 one extends original phase space to some larger one. In this larger space everything is 
 normal, motion is like canonical transformation in classical
  mechanics or unitary transformation in quantum theory. Then we return back to our physical
   reality, projecting this enlarged prereality and thus creating
 some probabilities [2].This was a general road of constructing 
 quantum mechanics and measurement theory, relativistic 
 quantum theory.\\
It is intriguing, that two milestones on this road: Polyakov theory of strings [11] 
and Parisi's theory of replica symmetry breaking [12] has been done almost simultaneously.
\\
We hope, that it should give a solution to other complex situations
as well. As it is impossible to understand human without soul,
it is impossible also to understand other complex systems without
 prereality (considering only observable) to define state of complex system. \\
 The degree of complexity of prereality could correspond to the degree of height on the modality
 staircase.\\ 
{\bf 6. Random Energy Model (REM) as a paradigm of complex systems.}
There are universality classes according to reflection schemes and information processing 
mechanism (language). If physical system has logarithmic free energy and local symmetric 
property as a mechanism for information processing, {\bf it should resemble some phase 
of REM}. REM version is a compressed form (some modalities are missed) of the theory,
its essence.  For other languages with nontrivial grammar there should be other universality 
classes, where free energy like variable scales with the number of degrees in a sublinear way [13]. \\ 
{\bf 7. Emergence during the reflections.}
 We interact with environment with our actions (magnetizations, phenotype),originated 
 via some general principles (ideal, genotype, ferromagnetic type coupling). \\
{\bf 8. Existence of one observable total probability in reality (Jaynes)}[1].\\
It is very hard to find  vacuum distributions and distribution of variables in a single vacuum 
in spin-glasses but we can calculate observable. 

To model complex systems one first should clarify the number
and type of reflections and the degree of harmony, as well 
as the suppression of local symmetries in  physical systems.
Then one should decide how to expand configuration space to 
construct proper preality.\\
It is interesting to analyze stock market under this point of view. We see  the following chain
of reflections: goods-money-stocks-derivatives. It is even longer than in strings [2]. 
If we apply our "the more is complex the system, the higher it should climb on modality staircase"  principle,  it will 
be reasonable to look for quantum mechanics like expanded prereality as a natural phase space here. 
There are some indications [14].\\
This work was supported by ISTC fund Grant A-102.

\end{document}